\documentclass[preprint,5p,twocolumn,amssymb,amsmath,amsfonts,bbold]{elsarticle}
\usepackage{graphicx}
\usepackage{amsmath}
\usepackage{amsfonts}
\usepackage{amssymb}
\usepackage{bbm}
\usepackage{xcolor}

\usepackage[normalem]{ulem}

\begin{document}
\title{Chiral Magnetic Effect in the Dirac-Heisenberg-Wigner formalism}
\author[wigner_rmi]{D\'aniel Ber\'enyi\corref{cor1}}
\cortext[cor1]{Corresponding author, berenyi.daniel@wigner.mta.hu}
\author[wigner_rmi]{P\'eter L\'evai}
\address[wigner_rmi]{Wigner RCP, Institute for Particle and Nuclear Physics, P.O. Box 49, Budapest 1525, Hungary}

\begin{abstract}
	Abstract: In this paper the emergence of the Chiral Magnetic Effect (CME) and the related anomalous current is investigated using the real time Dirac-Heisenberg-Wigner formalism. This method is widely used for describing strong field physics and QED vacuum tunneling phenomena as well as pair production in heavy-ion collisions. We extend earlier investigations of the CME in constant flux tube configuration by considering time dependent electric and magnetic fields. In this model we can follow the formation of axial charge separation, formation of axial current and then the emergence of the anomalous electric current. Qualitative results are shown for special field configurations that help to interpret the predictions of CME related effects in heavy-ion collisions at the RHIC Beam Energy Scan program.
\end{abstract}

\maketitle
\section{Introduction}

Quantum chromodynamics (QCD) comes with the intruiging phenomenon of topologically charged field configurations that presumed to develop charge separation in the presence of background magnetic fields \cite{KharzeevTopoChargeHIC, Fukushima:2008xe}. This process is known as the Chiral Magnetic Effect (CME) and the electric current that is the result of this charge separation might be an experimentally verifiable consequence of the theory. With particle accelerators like the Relativistic Heavy-Ion Collider (RHIC) and the Large Hadron Collider (LHC) covering ever larger range of the collision energy parameter space, enough experimental data have been accumulated to study QCD at these energy frontiers. The color electric and color magnetic fields forming at short initial times are so strong that they reach the critical fields strengths of $E_{cr} = B_{cr} = \frac{m^2 c^3}{g\hbar}$. Meanwhile the bypass of highly charged nuclei in non-central collisions near the speed of light induces extremely strong electrodynamical magnetic fields. The colliding nuclei release quarks and antiquarks to form a plasma exposing them to the very strong magnetic background field that starts to polarize them, align their spins (helicities) with their momentum and separate the quarks and anti-quarks based on their charges. The color fields in these processes are often modeled by a color flux tube that is the simplest non-trivial topologically charged field configuration. Since such topologically charged fields produce chirality imbalance, the difference in quark numbers manifests in charge difference and thus electric current. This potential observability sparked interest in studying the chiral magnetic effect from different perspectives.

Many studies investigated the effect using different methods, such as real-time lattice simulations \cite{BuividovichRealTimeLattice, MullerRealTimeLattice, TanjiAnomChargeLattice}, including backreaction \cite{HebenstreitCME} or hydrodynamics \cite{Yee:2013cya} where the anomaly gives rise to the Chiral Magnetic Wave. The strong field based description propose a natural, dynamical microscopical process that can account for the effect at least in the simplest topologically charged configuration: the flux tube. Initial studies in this area were done by Ref. \cite{FukushimaRealTimeCME} considering a constant Abelian fluxtube. A natural extension of the constant flux tube model relevant to heavy-ion collisions is to take into account the temporal change of the external and color fields. A motivation for doing so is that one may expect a better understanding of the time evolution of the system by invoking models from the semi-classical strong field description that already proven valuable in the case of hadron spectra and particle production description in heavy-ion collisions \cite{SkokovLevaipTspectra, SkokovLevaiSU2spectra, LevaiSkokovSU2heavyq}. Along these lines in this work we use a real-time Wigner function based description known as the Dirac-Heisenberg-Wigner formalism to model the quark fields under the influence of homogeneous but time dependent external fields. By restricting the model to homogeneous fields, we can also avoid the difficulties pointed out in Ref. \cite{WuWignerCMEmuproblems}. 

First, we review the fundamentals of the Wigner-function based description in Section 2. Then in Section 3. we study a simple toy field configuration to assess qualitatively the description of the chiral magnetic effect and the resulting currents. Next, we turn to more realistic field configurations aiming at mimicing the ones that are expected to be formed in relativistic heavy-ion collosions at RHIC.

\section{Theoretical background}
The Dirac-Heisenberg-Wigner (DHW) formalism is a real-time description of the one-particle Wigner function to describe the spatio-temporal evolution of a fermionic field under the influence of classical external fields. The choice of such a classical model is justified on one hand by the expected criticality of the gluonic fields in relativistic heavy-ion collisions and by the extremity of the magnetic background as detailed in the Introduction. This method has proven useful in studying the interplay of electromagnetic fields with spatio- temporal variability \cite{KohlfurstWigner1, KohlfurstWigner2}.

The DHW equations can be formulated for U(1) and SU(N) fields, the latter being much more complicated in terms of structure. However, it was shown \cite{FukushimaRealTimeCME, LevaiSkokovSU2heavyq}, that due to the strong Abelian dominance many aspects can be readily reproduced by the much simpler U(1) description because in that case the color fields can be diagonalized and decoupled to multiple copies of the equivalent Quantum Electrodynamics (QED) theory. This in turn was already widely investigated in the context of vacuum structure and pair production since the original formulation of the description by \cite{BirulaDHW}.

The U(1) Wigner function $W(\vec{x}, \vec{p}, t)$ of a relativistic particle with mass $m$ and charge $g$ can be expanded on the Dirac spinor basis:
	
\begin{equation}
W(\vec{x}, \vec{p}, t) = \frac{1}{4}\left[ 1\mathbbm{s} + i\gamma_5\mathbbm{p} + \gamma^{\mu}\mathbbm{v}_{\mu} + \gamma^{\mu}\gamma_5\mathbbm{a}_{\mu} + \sigma^{\mu\nu}\mathbbm{t}_{\mu\nu}\right]
\end{equation}
such that the components represent scalar, pseudoscalar, vector, axial-vector and tensor quantities respectively. For the latter, we introduce the following three component vectos: $(\mathbbm{\vec{t}}_1)_i = \mathbbm{t}_{0i} - \mathbbm{t}_{i0}$ and $(\mathbbm{\vec{t}}_2)_i = \epsilon_{ijk} \mathbbm{t}_{jk}$. For homogeneous fields this results in a partial differential equation system of 16 real components, that following \cite{AlkoferIDHW} reads:

\begin{align}
&D_t \mathbbm{s}				& & & &    & &-& & 2\vec{p} \cdot \mathbbm{\vec{t}}_1         &=& 0\;,& \label{EqDHWStart} \\
&D_t \mathbbm{p}				& & & &    & &+& & 2\vec{p} \cdot \mathbbm{\vec{t}}_2         &=& 2m\mathbbm{a}_0\;,& \\
&D_t \mathbbm{v}_0			& &+& & \vec{D}_{\vec{x}} \cdot \vec{\mathbbm{v}} & & & &             &=& 0\;,& \\
&D_t \mathbbm{a}_0			& &+& & \vec{D}_{\vec{x}} \cdot \vec{\mathbbm{a}} & & & &             &=& 2m\mathbbm{p}\;,& \\
&D_t \mathbbm{\vec{v}}		& &+& & \vec{D}_{\vec{x}} \mathbbm{v}_0 & &+& & 2\vec{p}\times \vec{\mathbbm{a}}  &=& -2m\vec{\mathbbm{t}}_1\;,& \\
&D_t \mathbbm{\vec{a}}		& &+& & \vec{D}_{\vec{x}} \mathbbm{a}_0 & &+& & 2\vec{p}\times \vec{\mathbbm{v}}  &=& 0\;,& \\
&D_t \mathbbm{\vec{t}}_1	& &+& & \vec{D}_{\vec{x}} \times \mathbbm{\vec{t}}_2 & &+& & 2\vec{p}\mathbbm{s}  &=& 2m\vec{\mathbbm{v}}\;,& \\
&D_t \mathbbm{\vec{t}}_2	& &-& & \vec{D}_{\vec{x}} \times \mathbbm{\vec{t}}_1 & &-& & 2\vec{p}\mathbbm{p}  &=& 0\;.& \label{EqDHWEnd}
\end{align}
where the evolution operators are given without any approximations by:

\begin{eqnarray}
              D_t &=& \partial_t + g\vec{E} \cdot \vec{\nabla}_{\vec{p}}\;,\\ 
\vec{D}_{\vec{x}} &=& g\vec{B} \times \vec{\nabla}_{\vec{p}}\;.
\end{eqnarray}

The components that are of interest to us are the current density $\mathbbm{\vec{v}}$, the axial current density $\mathbbm{\vec{a}}$ and the axial charge density $\mathbbm{a}_0$.

The initial conditions for vacuum are only non-vanishing for the mass density and the current density:
\begin{eqnarray}
     \mathbbm{s} (\vec{p}, t=-\infty) &=& -\frac{2m}{\omega(\vec{p})}\;,\\ 
\vec{\mathbbm{v}}(\vec{p}, t=-\infty) &=& -\frac{2\vec{p}}{\omega(\vec{p})}\;,
\end{eqnarray}
where $\omega^2(\vec{p})=m^2+p_x^2+p_y^2+p_z^2$.

As we are mainly interested in light quark production, we take the $m\rightarrow0$ limit. This results in only 8 coupled equations:

\begin{align}
&D_t \mathbbm{v}_0& &+& &\vec{D}_{\vec{x}} \cdot \vec{\mathbbm{v}}& & & &= 0\;,& \\
&D_t \mathbbm{a}_0& &+& &\vec{D}_{\vec{x}} \cdot \vec{\mathbbm{a}}& & & &= 0\;,& \\
&D_t \mathbbm{\vec{v}}& &+& &\vec{D}_{\vec{x}} \mathbbm{v}_0& &+& 2\vec{p}\times \mathbbm{\vec{a}} &= 0\;,\\
&D_t \mathbbm{\vec{a}}& &+& &\vec{D}_{\vec{x}} \mathbbm{a}_0& &+& 2\vec{p}\times \mathbbm{\vec{v}} &= 0\;.&
\end{align}

To further simplify the equations we use the Method of Characteristics \cite{AlkoferIDHW}. We integrate the electric field to obtain the vector potential, and use that to shift the momentum variable $\tilde{\vec{p}} = \vec{p} + g \int \vec{E}(t) dt$ to get rid of the $g\vec{E} \cdot \vec{\nabla}_{\vec{p}}$ term in $D_t$. This way only those momentum derivates remain that are multiplied by the magnetic field in $\vec{D}_{\vec{x}}$.

We use a global pseudo-spectral collocation in the three dimensional momentum space on Rational Chebyshev polynomials \cite{BoydRatCheb1, BoydRatCheb2}, and a 4th order explicit Runge-Kutta stepper in time. The numerical code is powered by OpenCL to utilize Graphical Processing Units (GPUs) for the dense tensor operations, that results in a factor of 30 speedup w.r.t. a conventional implementation.

The numerical solver was verified on the two important analythic solutions: the time dependent Sauter electric field case \cite{Kruglov} and the stationary magnetic field solution given in \cite{BirulaDHW}.

During the time evolution, we record the following momentum space integrals:
\begin{align}
&\mathbbm{v}^{\mu}(t) & &=& & \frac{1}{(2\pi)^3} \int\limits_{-\infty}^{\infty} {\rm d}p^3 \mathbbm{v}^{\mu}(t, \vec{p})\;,&\\
&\mathbbm{a}^{\mu}(t) & &=& & \frac{1}{(2\pi)^3} \int\limits_{-\infty}^{\infty} {\rm d}p^3 \mathbbm{a}^{\mu}(t, \vec{p})\;.&
\end{align}

These quantities are corresponding to the total electric charge and current as well as the total axial charge and current respectively. Electric charge is conserved, so $\mathbbm{v}_0(t) = 0$, but the axial charge develops a non-zero value, since it is related to the chiral imbalance: $\mathbbm{a}_0( t = +\infty) = N_R-N_L$.

As the functions of interest are discretized on the Rational Chebyshev basis, a matching spectrally convergent Clenshaw-Curtis quadrature can be used to calculate these integrals precisely \cite{BoydInfiniteIntervalQuadrature}.

\section{Sauter field configuration}

Before applying the DHW formulation to study realistic field configurations it is worth taking a look at the outcome of simpler cases and verify that the model predictions match the expected CME characteristics. We chose the Sauter field for this initial study, as this field is widely studied and understood in the pair production picture and can be used to build intuition on how the system behaves.

The Sauter field is given by: 

\begin{align}
f(t) = A \cosh^{-2}\left(\frac{t}{\tau}\right).
\label{fig_th_Sauter}
\end{align}

\begin{figure}
	\centerline{%
		\includegraphics[width=0.9\linewidth]{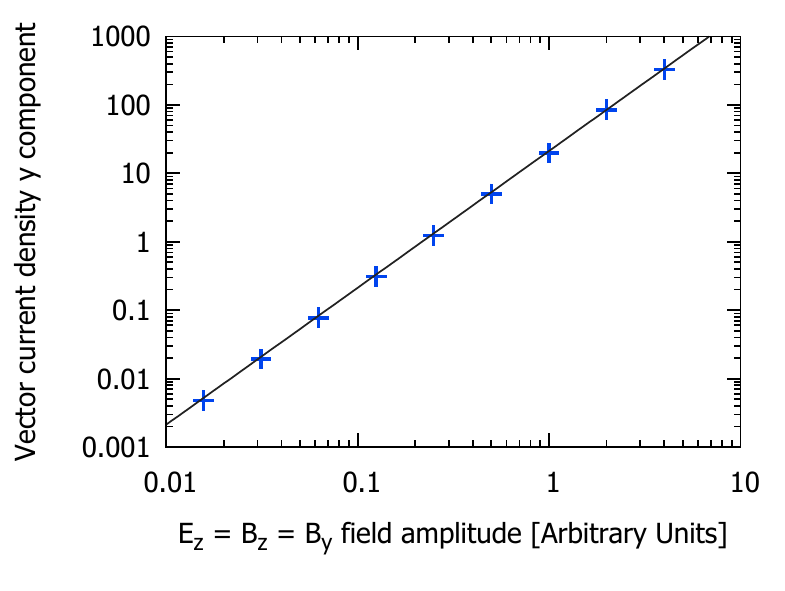}}
	\caption{ The anomalous ($\mathbbm{v}_y$) component of the vector current density for different field amplitudes $E_z = B_y = B_z = A$. The datapoints show an $A^2$ dependence (solid line). }
	\label{fig_th_AmplVy}
\end{figure}

In contrast to the massive case, where field amplitude scales are set by $m^2/g$, in the massless limit there is no such specific intrinsic scale. But when we choose one (e.g. to minimize the cost of the numerical computation) it also sets the time scale $\tau_0 = 1 / \sqrt{g A}$. If we let $E_z(t) = B_z(t) = B_y(t) = f(t)$ and calculate the anomalous electric current density we find an $A^2$ dependence as shown on Figure \ref{fig_th_AmplVy}. in agreement with the Schwinger formula for pair production, or in the CME setting with Ref. \cite{FukushimaRealTimeCME}

\begin{figure}
	\centerline{%
		\includegraphics[width=0.9\linewidth]{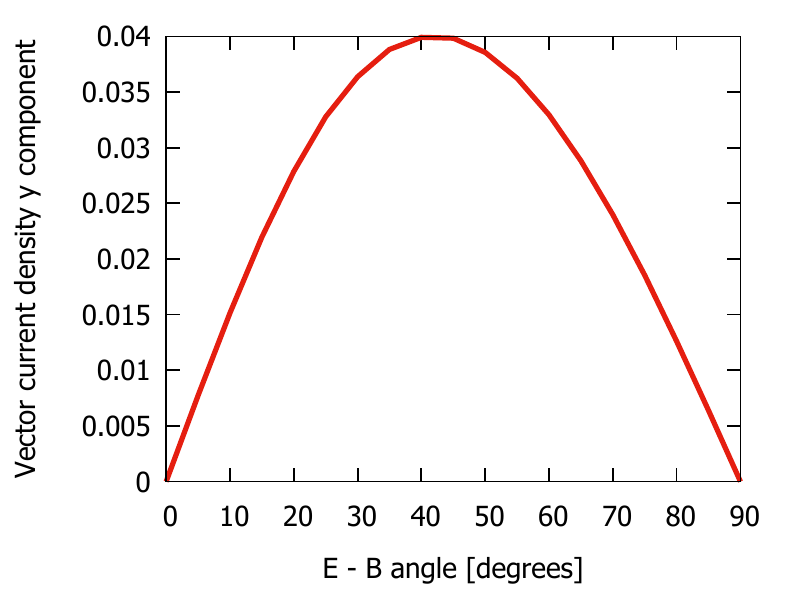}}
	\caption{ The change of the asymptotic ($t\rightarrow\infty$) CME current ($\mathbbm{v}_y$) with the angle between the electric and magnetic fields.}
	\label{fig_th_ang}
\end{figure}

The next important thing in the CME picture is to verify that the anomaly effect only exists when none of the three fields, $E_z$, $B_z$ and $B_y$ vanish. It is again well known in the context of pair production that magnetic fields alone create no observable currents, so it is enough to investigate the dependence of the CME effect on the angle between a fixed non-zero electric field and a magnetic field. Figure \ref{fig_th_ang}. shows that the CME current decays linearly to zero when the magnetic field is either becoming parallel or perpendicular to the electric field.

\begin{figure}
	\centerline{%
		\includegraphics[width=0.9\linewidth]{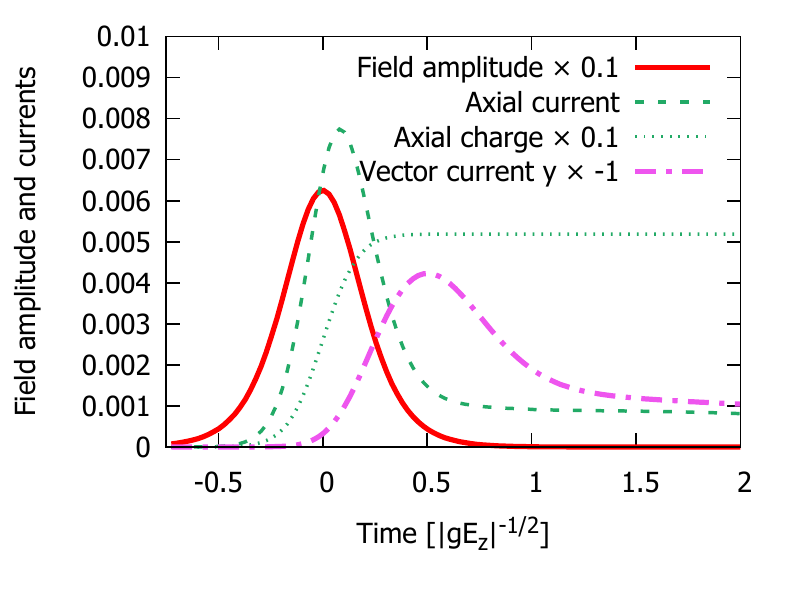}}
	\caption{Time dependence of the external field, the axial current density, the axial charge density and the anomalous vector current density.}
	\label{fig_th_timedep}
\end{figure}

It is instructive to take a look at the time evolution of the fields on Figure \ref{fig_th_timedep}, because it clearly shows the chain of events that lead to the CME current. First, the external fields build up. This drives the formation of an axial current, followed by an axial charge separation, that results in the charge displacement that creates the vector current. As the external driving fields decay, the induced charge and currents stabilize and attain their asypmtotic values, as in this formulation there is no back reaction involved.

\begin{figure}
	\centerline{%
		\includegraphics[width=0.9\linewidth]{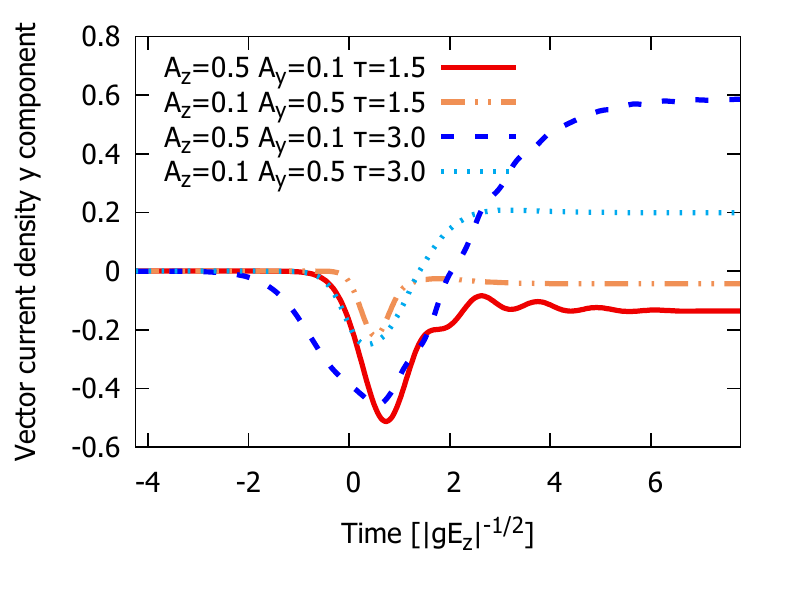}}
	\caption{Time dependence of the CME current density ($\mathbbm{v}_y$) for different amplitude and time scales.}
	\label{fig_th_amptaudep}
\end{figure}

Finally, we compare the interplay of temporal and amplitude scales by considering Sauter fields with different parameters. Let $E_z = B_z = A_z$ and $B_y = A_y$ be the amplitude of the Sauter fields of $E_z(t), B_z(t)$ and $B_y(t)$ respectively and let $\tau$ be the timescale of all of them (c.f. eq. \ref{fig_th_Sauter}). Figure \ref{fig_th_amptaudep}. shows the anomalous current for two time scales $\tau=1.5$ and $3$ and for two amplitude pairs, one of where the z direction is stronger $A_z=0.5; A_y=0.1$ and another where the y direction is stronger $A_z=0.1; A_y=0.5$. The results show, that for shorter Sauter fields, the $\mathbbm{v}_y$ current does not change sign while for longer fields a zero crossing occurs after the peak of the external fields. The relative amplitude of the fields only moderately scales this behavior, but does not enough alone to qualitatively change it. 

In strong field physics, there is a known qualitative change in the behaviour of the pair production process for the unitary value of the Keldysh parameter $\gamma = \frac{m}{eE\tau}$ that is defined by the ratio of the energy scales of the tunnelling particle and the external field (or the equivalent time scales) and this marks the transition between the perturbative and non-perturbative regimes \cite{GelisTanji_SchwingerReview}. In our case, there are also two competing scales set by $\tau_B = 1 / \sqrt{e B_y}$ and $\tau_E = 1 / \sqrt{g E_z}$ and we argue that the magnitude of the final CME vector current is highly dependent on the interplay of these timescales similarly to the Keldysh case.

\section{RHIC Beam Energy Scan Results}

\begin{figure}
	\centerline{%
		\includegraphics[width=1\linewidth]{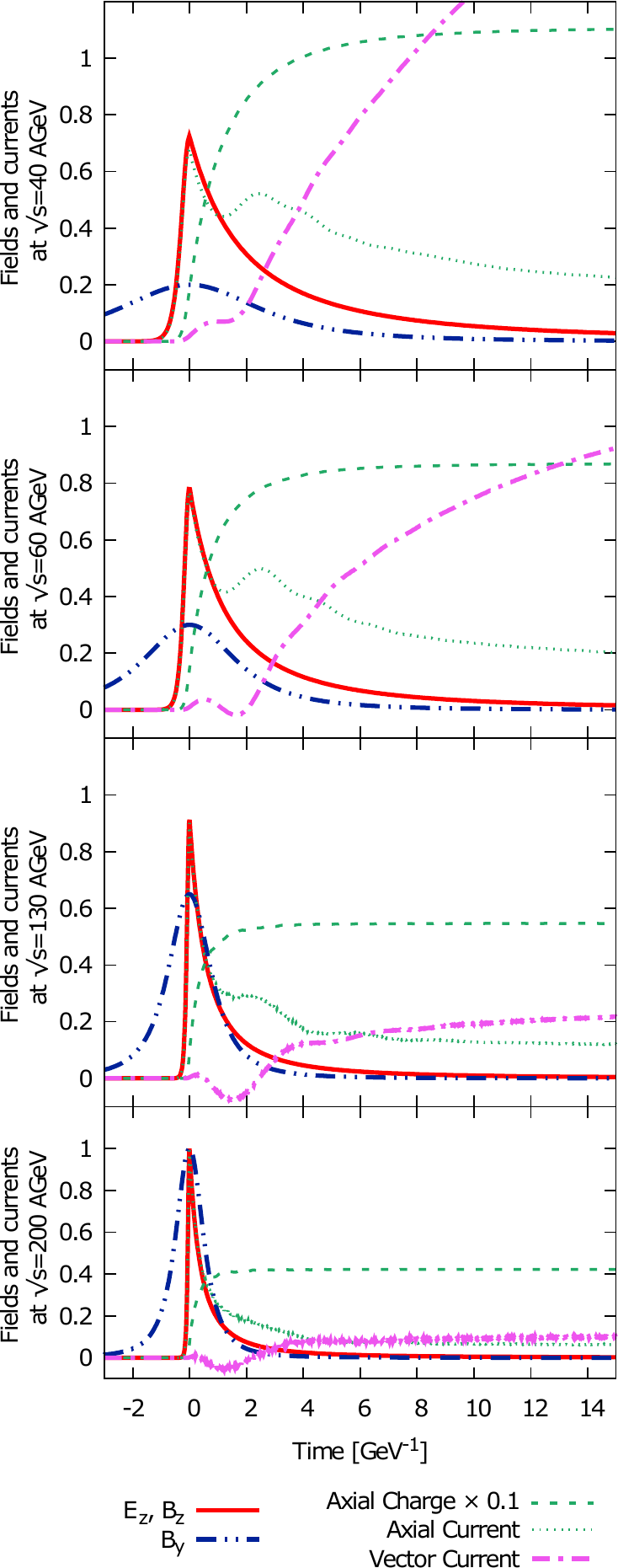}}
	\caption{Time dependence of the external field, the axial current, the axial charge and the vector current at collision energies $\sqrt{s} = 40, 60, 130$ and $200$ AGeV.}
	\label{fig_rhicexp_timedep}
\end{figure}

As the DHW equations are using prescribed field configurations for the E and B fields, we need a model of these that mimic heavy-ion collisions. We start with the following time dependent function 

\begin{equation}
\Phi(t, \tau, A, \kappa) = A \cdot
\left\{
\begin{aligned}
&\rm{cosh}^{-2}(10 t/\tau) & & t < 0,\\
&(1+t/\tau)^{-\kappa} & & t \ge 0,
\end{aligned}
\right. \label{eq_Phi}
\end{equation}

where we set $\kappa = 2$ and model the external fields in heavy-ion collisions as follows (\cite{SkokovLevaipTspectra} eq. 26. for $E_z, B_z$ and \cite{ZakharovEMRQGP} eq. 12. for $B_y$):

\begin{align}
&e\vec{E} &=& \{0, 0,                                & \Phi(t, \tau, A_{Ez}, \kappa) \}  \;,&\\
&e\vec{B} &=& \{0, A_{By}\left(1+\frac{t^2}{\tau^2}\right)^{-3/2}, & \Phi(t, \tau, A_{Bz}, \kappa)  \} \;.&
\end{align}

 and the other quantities are related to the center of mass energy $\sqrt{s}$ as follows \cite{LevaiSkokovSU2heavyq}:

\begin{align}
& \tau & &=& &\frac{0.75}{Q_s} \frac{\sqrt{s}_{\rm{RHIC}}}{\sqrt{s}} \;,& \label{EqTau}\\
& A_{Ez} = A_{Bz} & &=& &Q_s^{2} \left( \frac{\sqrt{s}}{\sqrt{s}_{\rm{RHIC}}} \right)^{\lambda}\;,&\\
& A_{By} & &=& &0.2 Q_s^{2} \frac{\sqrt{s}}{\sqrt{s}_{\rm{RHIC}}} \;
\end{align}

with $Q_s = 1 \rm{GeV}$ and $\sqrt{s}_{\rm{RHIC}} = 200 \rm{AGeV}$. The $z$ amplitude scale $\lambda$, is the gluon saturation scale, that is chosen to be 0.2 in accordance with the literature \cite{Golec_Sat, Kowalski_Sat}.

Following the above scaling rules the tabulated values of the relevant quantities for some representative values of the RHIC beam energy scan program are the following:

\begin{center}
	\begin{tabular}{ r | c | c | c | c | c }
		\hline
		$\sqrt{s} [\rm{AGeV}]$ & 200 & 130 & 60 & 40\\ \hline \hline
		$A_{Ez}  [\rm{GeV}^2]$ & 1.00 & 0.92 & 0.79 & 0.73\\ \hline
		$\tau [\rm{GeV}^{-1}]$ & 0.75 & 1.15 & 2.5 & 3.75\\ \hline
		
	\end{tabular}
\end{center}

Figure \ref{fig_rhicexp_timedep}. shows the shape of the external fields, and the time evolution of the x component of the axial current, the axial charge and the y component of the vector current for different collision energies. The simulation results outline the following picture: As seen from eq. (\ref{EqTau}) smaller energies result in larger $\tau$, which is known in the strong field picture to increase the magnitude of currents and make the temporal dynamics last longer. This is clearly observed in the time dependence of the $A_x$, but even more pronounced in the anomalous electric current, $V_y$, which slowly approaches its asymptotic value, since the driving fields have only a cut-power law decay (c.f. eq. \ref{eq_Phi}).


The unexpected phenomena at almost all energies is that the anomalous current starts with a negative sign and undergoes a reversal at $t \approx 2 \ {\rm GeV^{-1}}$
after the collision. Between 40-60 AGeV the negative dip disappears that we attribute to the larger width of the pulse. Precise values depend on the external field models, its amplitude, gradients, decay rate, but for pulse like fields the overall behavior is the same.

\section{Discussion}
In this work we investigated the Chiral Magnetic Effect in the Dirac-Heisenberg-Wigner formalism in case of time dependent homogeneous fields realized inside a fluxtube. Our model has reproduced the basic features of the onset of anomalous current which is expected from the presence of the strong magnetic fields in heavy-ion collisions. After understanding the effect in the Sauter field configuration we investigated the predictions for a more realistic model field.

We have applied proper parametrization for different RHIC collision energies in the 40-200 AGeV range and performed our analysis. We obtained the following results. First, our calculations indicate the expected disappearance of Chiral Magnetic Effect at energies above 200 AGeV. Second, above 60 AGeV we have found an interesting sign change in the anomalous current. We could understand from our model that both effects are expected to be connected to the temporal extension of the fields and less sensitive to other field characteristics. The reason is the low exponent in the amplitude scale given by the gluon saturation scale $\lambda$ that makes the field amplitudes change much less with $\sqrt{s}$ than the time scales. Very high energies create very short time scales that limits formation of all current. For lower energies the time scales are longer, and at one point the process transitions to a different regime and this transition is responsible for the rapid increase of the anomalous current.




\section*{Acknowledgements}
The authors thank to Vladimir V. Skokov for discussions and commenting the manuscript. Parts of this work was done during the visit of D.B. at BNL and would like to thank the hospitality and intriguing discussions. The simulations were run on the Wigner GPU Lab AMD Radeon GPU cluster. Special thanks to M\'at\'e Ferenc Nagy-Egri for cluster technical support.
This work was supported in part by the Hungarian National Bureau for Research, Development and Innovation, NKFIH Grants No. NK106119, No. K16-120660, No. K123815 and TeT12-CN-1-2012-0016


\begin{thebibliography}{1}
	
	\bibitem{KharzeevTopoChargeHIC}
	D. E. Kharzeev, L. D. McLerran, H. J. Warringa
	Nucl. Phys. {\bf A803} (2008) 227-253.
	
	\bibitem{Fukushima:2008xe}
	K.~Fukushima, D.~E.~Kharzeev and H.~J.~Warringa,
	Phys.\ Rev.\ D {\bf 78} (2008) 074033
	doi:10.1103/PhysRevD.78.074033
	
	\bibitem{BirulaDHW}
	I. Bialynicki-Birula, P. Gornicki, J. Rafelski,
	Phys. Rev. {\bf D44 } (1991) 1825-1835.
	
	\bibitem{AlkoferIDHW}
	F. Hebenstreit, R. Alkofer, H. Gies,
	Phys. Rev. {\bf D82 } (2010) 105026.
	
	\bibitem{FukushimaRealTimeCME}
	K. Fukushima, D. E. Kharzeev, H. J. Warringa
	Phys. Rev. Lett. {\bf 104 } (2010) 212001.
	
	\bibitem{ZakharovEMRQGP}
	B.G. Zakharov
	Phys. Lett. {\bf B737} (2014) 262-266.
	
	\bibitem{WuWignerCMEmuproblems}
	Y. Wu, D-f. Hou, H-c. Ren
	Phys. Rev. {\bf D 96}, 096015 (2017)
	
	\bibitem{SkokovLevaipTspectra}
	V. V. Skokov, P. L\'evai
	Phys. Rev. {\bf D71} (2005) 094010.
	
	\bibitem{SkokovLevaiSU2spectra}
	V. V. Skokov, P. L\'evai
	Phys. Rev. {\bf D78} (2008) 054004.
	
	\bibitem{LevaiSkokovSU2heavyq}
	P. L\'evai, V. V. Skokov
	Phys. Rev. {\bf D82} (2009) 074014.
	
	\bibitem{TanjiAnomChargeLattice}
	N. Tanji, N. Mueller, J. Berges
	Phys. Rev. {\bf D93} (2016) 074507.
	
	\bibitem{MullerRealTimeLattice}
	N. M\"uller, S. Schlichting, S. Sharma
	Phys. Rev. Lett. {\bf 117}, (2016) 142301. 
	
	\bibitem{BuividovichRealTimeLattice}
	P. V. Buividovich, S. N. Valgushev
	PoS LATTICE2016 (2016) 253
	
	
	\bibitem{Yee:2013cya} 
	H.~U.~Yee and Y.~Yin,
	Phys.\ Rev.\ C {\bf 89}, no. 4, 044909 (2014)
	doi:10.1103/PhysRevC.89.044909
	
	\bibitem{Kruglov}
	S. I. Kruglov,
	Radiat. Phys. Chem. {\bf 75} (2006) 723-728.
	
	\bibitem{BoydRatCheb1}
	J. P. Boyd, 
	Computers and Mathematics with Applications 41. (2001) 1293-1315.
	
	\bibitem{BoydRatCheb2}
	J. P. Boyd, 
	J. Comp. Phys. {\bf 69(1)} (1987) 112-142. 
	
	\bibitem{BoydInfiniteIntervalQuadrature}
	J. P. Boyd, 
	J. Sci. Comp. Vol. 2 No. 2 (1987)
	
	\bibitem{Golec_Sat}
	K. Golec-Biernat and M. Wüsthoff,
	Phys. Rev. {\bf D59}, 014017, (1998)
	
	\bibitem{Kowalski_Sat}
	H. Kowalski, L. Motyka and G. Watt,
	Phys. Rev. {\bf D74}, 074016 (2006)
	
	\bibitem{GelisTanji_SchwingerReview}
	F. Gelis, N. Tanji, 
	Progress in Particle and Nuclear Physics, Vol. 87 pp. 1-49, (2016)
	
	
	\bibitem{KohlfurstWigner1}
	C. Kohlf\"urst, R. Alkofer, 
	Phys. Lett. {\bf B 756} (2016) pp. 371-375.
	
	\bibitem{KohlfurstWigner2}
	C. Kohlf\"urst, R. Alkofer, 
	Phys. Rev. {\bf D97}, (2018) 036026 

	\bibitem{HebenstreitCME}
	N. Mueller, F. Hebenstreit, J. Berges,
	Phys. Rev. Lett. {\bf 117} (2016) 061601
	
	
	
	
	
	
\end{thebibliography}
\end{document}